\documentclass[
    aps,
    pra,
    onecolumn,
    letterpaper,
    10pt,
    superscriptaddress,
    notitlepage,
    amsmath,
    amssymb,
    floatfix
]{revtex4-1} 

\usepackage{bm,enumerate,dcolumn,tikz,graphicx,color,amsmath,amssymb}
\usepackage{epstopdf}
\usepackage{empheq}
\usepackage{capt-of}
\usepackage{pgfplots}

\def\beq{\begin{equation}}
\def\eeq{\end{equation}}
\def\bmat{\begin{displaymath}}
\def\emat{\end{displaymath}}
\def\bear{\begin{eqnarray}}
\def\eear{\end{eqnarray}}
\def\ba{\begin{eqnarray}}
\def\ea{\end{eqnarray}}
\def\bery{\begin{array}}
\def\ery{\end{array}}
\def\bit{\begin{itemize}}
\def\eit{\end{itemize}}
\def\ben{\begin{enumerate}}
\def\een{\end{enumerate}}
\def\btab{\begin{tabular}}
\def\etab{\end{tabular}}
\def\btbl{\begin{table}}
\def\etbl{\end{table}}
\def\bfig{\begin{figure}[htb]}
\def\efig{\end{figure}}
\def\bpic{\begin{picture}}
\def\epic{\end{picture}}

\def\be{\begin{equation}}
\def\ee{\end{equation}}
\def\ba{\begin{eqnarray}}
\def\ea{\end{eqnarray}}

\begin{document}

\title{
Hunting for topological dark matter with atomic clocks
} 
\author{A. Derevianko}
\affiliation{Department of Physics, University of Nevada, Reno, Nevada 89557, USA}
\author{M. Pospelov}
\affiliation{Department of Physics and Astronomy, University of Victoria, Victoria, British Columbia V8P 1A1, Canada}
\affiliation{Perimeter Institute for Theoretical Physics, Waterloo, Ontario N2J 2W9, 
Canada}

\maketitle

The cosmological applications of atomic clocks~\cite{ChoHumKoe10,BloNicWil14,Lev99} so far have been limited to searches of 
the uniform-in-time drift of fundamental constants~\cite{RosHumSch08}. 
We point out that a transient in time change of fundamental 
constants can be induced by dark matter objects that have large spatial extent, such as stable topological defects~\cite{Vilenkin:1984ib} built from light non-Standard Model fields. 
Networks 
of correlated atomic clocks, some of them already in existence~\cite{Predehl2012}, such as the Global Positioning System, 
can be used as a powerful tool to search for the 
topological defect dark matter, thus providing 
another important fundamental physics application to the 
ever-improving accuracy of atomic clocks. During the encounter with an extended dark matter object, 
as it sweeps through the network,
initially synchronized clocks will become desynchronized.  Time discrepancies between spatially-separated clocks are expected to exhibit a distinct signature, encoding defect's space structure and  its interaction strength with 
atoms.

Despite solid evidence for the existence of dark matter ($\sim 25\%$ of the global 
energy budget in the Universe, and $\rho_{\rm DM} \simeq 0.3$ GeV$/$cm$^3$ in
Solar system neighborhood \cite{RPP-2012}), its 
relation to particles and fields of the Standard Model (SM) remains a mystery. 
While searches of particle dark matter (DM) are being actively pursued \cite{Bertone:2010zza},  
there is also significant interest 
to alternatives, among which the DM composed from very light fields.
Depending on the initial field configuration 
at early cosmological times, such light fields 
could lead to dark matter via coherent oscillations around the minimum of their potential, and/or form non-trivial stable field configurations in physical 3D space if their potential allows for such possibility. This latter option, 
that we will generically refer to as the topological defects (TD), is the main interest of our paper.  The light masses of fields forming the TDs could lead to a large, indeed macroscopic, size for a 
defect. Their encounters with the Earth, combined together with the DM-SM coupling, 
can lead to novel signatures of dark matter expressed generically in terms of the ``transient effects''.
These effects, coherent on the scale of individual detectors, are temporary shifts in frequencies
and phases of measuring devices, rather than large energy depositions as is the case for microscopic DM. 
In this paper we suggest the possibility of a new search technique for the topological defect dark matter (TDM), based on 
a network of atomic clocks. 

Atomic clocks are arguably the most accurate scientific instruments ever build, 
reaching the $10^{-18}$ fractional inaccuracy~\cite{ChoHumKoe10,BloNicWil14}.
 Attaining this accuracy requires that the quantum oscillator be well protected from environmental noise and perturbations well controlled and characterized. This opens intriguing prospects of using clocks to study subtle effects, and it is natural to ask if such accuracy can be harnessed for dark matter searches. 

To put our discussion on concrete grounds, we introduce a collection of light fields  beyond the SM, that can form 
TDs of different dimensionality: monopoles (0d), strings (1d), and domain walls (2d). Exact nature of such defects 
depends on the composition of the dark sector, and on self-interaction potential~\cite{Vilenkin:1984ib}. 
For this paper we  take a simplified approach,
and call $\phi$ a generic light field from the dark sector, would it be scalar or vector, that forms a network of TD at some 
early stage of cosmological history.
The transverse size of the defect is determined by the field Compton 
wavelength $d$, that is in inverse relation to the typical mass scale of the light fields, 
$d\sim \hbar/( m_\phi c)$. The fields we are interested in are ultralight: for an Earth-sized defect, the mass scale is $10^{-14}\, \mathrm{eV}$.
In our simplified approach we capture only gross features of TDs~\cite{Vilenkin:1984ib}, and call $A$ 
the amplitude of the field change inside and outside a TD, $A = \phi_{\rm inside} - \phi_{\rm outside} $,
 also choosing the outside value of the field to be zero.

The energy density of TDM averaged over large number of defects is controlled by  the energy density inside the defect, $\rho_{\rm inside} \sim A^2/d^2$, and 
the average distance between the defects,  $L$, through natural scaling relation: 
\begin{equation}
\label{rtdm}
\rho_{\rm TDM} \sim  \rho_{\rm inside} d^{3-n}  L^{n-3} (\hbar c)^{-1} \sim A^2d^{1-n}L^{n-3} (\hbar c)^{-1},
\end{equation} 
where 
$n=0,1,2$ for a monopole, string or a domain wall, and we measure $A$ in units of energy.

Right combination of parameters can give a 
significant contribution to or even saturate $\rho_{\rm DM}$. The average time between ``close encounters'' 
with TD, $r\leq d$, is set by the galactic velocity of such objects $v_g$,
\begin{equation}
\label{T}
{\cal T}\simeq  \frac{1}{v_g}\times \frac{L^{3-n}}{d^{2-n}} = \frac{1}{v_g}\times \frac{A^2}{\rho_{\rm TDM}d}\times\frac{1}{\hbar c}.
\end{equation} 
Velocity of galactic objects around the Solar system is an input parameter that is relatively well known, 
and for the purpose of estimates one can take $v_g \simeq 10^{-3} \times c \approx 300 \, \mathrm{km/s}$. If  the parameter ${\cal T}$ is on the order of few years
or less, then it is reasonable to think of a detection scheme for
TD crossing events. 

The most crucial question is how the fields forming the defect interact with the SM. 
All possible types of interaction between TD and SM fields can be classified using the so-called ``portals'', 
the collection of gauge-invariant operators of the SM coupled with the operators from the dark sector \cite{Essig:2013lka}. 
Throughout the rest of  this paper, we are going to be interested in a more general form of the SM-TD interaction in the form of the 
quadratic scalar portal, 
\begin{eqnarray}
\label{interaction}
-{\cal L}_{\rm int}=  \phi^2\left(\frac{m_e \bar \psi_e \psi_e  }{\Lambda^2_e}  + \frac{m_p \bar \psi_p \psi_p  }{\Lambda^2_p}  -\frac{1}{4\Lambda_\gamma^2} 
F_{\mu\nu}^2 + ...\right)\\
\nonumber
\rightarrow ~~m_{e,p}^{\rm eff} = m_{e,p}\left(1+ \frac{\phi^2}{\Lambda_{e,p}^2}\right);~~\alpha^{\rm eff}= \frac{\alpha }{1-\phi^2/\Lambda_{\gamma}^2}
\end{eqnarray}
Since inside the TD, by assumption, $\phi^2 \rightarrow A^2$ and outside $\phi^2 \rightarrow 0$ 
this portal renormalizes masses and couplings only when the TD core overlaps with the quantum device.
Here $m_{e,p}$ and $\psi_{e,p}$ are electron and proton masses and fields, and $F_{\mu \nu}$ are electromagnetic tensor components.
The appearance of  high-energy scales $\Lambda_X$ in the denominators of (\ref{interaction})
signifies the effective nature of these operators, implying that at these scales the scalar portals will be replaced by some 
unspecified fundamental theory (the same way as electroweak theory of the SM replaces effective four-fermion 
weak interaction  at the electroweak scale). 
The SM field dependence in (\ref{interaction})  replicates corresponding pieces from the SM sector Lagrangian density,
and this leads to the identification (the second line of Eq.~ (\ref{interaction})) of how masses  and the fine-structure constant $\alpha$ are modulated by the TD. 
Thus, for every coupling constant 
and SM particle mass scale $X$ one has to first order in $\phi^2$
\begin{equation}
\frac{\delta X}{X} = \frac{\phi^2}{\Lambda_X^2}.
\end{equation} 
Quadratic (as opposed to linear) dependence on $\phi$ leads to 
weakening of constraints imposed by 
precision tests of gravitational interactions \cite{Olive:2007aj}.
Both direct laboratory and astrophysical constraints on $\Lambda_X$ do not exceed $\sim 10$ TeV.
Additional background information on TDM, the types of interaction with the SM, and plausible scenarios for its 
abundances are provided in the supplementary information (SI). In particular, we present an explicit 
example of the so-called Abrikosov-Neilsen-Olessen 
string defect \cite{Abrikosov:1956sx,Nielsen:1973cs},  
with an increased value of $\alpha$ inside 
its core.

The main 
consequence of the interaction (\ref{interaction}) is a temporary shift of all masses and frequencies inside
the TD.  
Thus, the signature we are proposing to search for is a transient variation of fundamental constants.
In the limit of  large $\tau$, when the size of a TD is on astronomical scales, the effect of (\ref{interaction}) becomes 
identical to variations of couplings and masses over time with $\dot \alpha \simeq $ const, in which case  all the existing terrestrial constraints immediately apply~\cite{RosHumSch08}. 
In addition, during the TD crossing there is a new force acting on massive bodies, giving a
transient signature that can be explored with sensitive graviometers. Also, there are other ways of coupling TD to SM
such as the so-called axionic portals, $\partial_\mu \phi /f_a \times J_\mu$, where 
$J_\mu$ is the axial-vector current. 
It would lead to a transient ``loss'' of rotational/Lorentz invariance, and can be 
searched for with sensitive atomic magnetometers~\cite{Pospelov:2012mt,Pustelny:2013rza}. By design, atomic clocks are   
less sensitive to the coupling to spin, and for that reason we concentrate on (\ref{interaction}).

Clocks tell time by counting number of oscillations and multiplying them by the predefined period of oscillations $1/(2\pi\omega_0)$, where 
$\omega_0$ is the fixed unperturbed clock frequency.  
Experimentally relevant quantity is the total  phase accumulated by the quantum oscillator,
$\phi_0(t) = \int_{0}^{t} \omega_0 dt'$; then apparently the device time reading is $\phi_0(t)/\omega_0$.  TD would shift the oscillator frequency and thereby affect the phase or the time reading, $\phi(t) = \int_{0}^{t} (\omega_0 + \delta \omega(t') ) dt'$, where  $\delta \omega(t')$ is the quantum oscillator frequency variation caused by TD.
We  parameterize  
 $\delta \omega(t) = g f(t)$, where $g \propto A^2/\Lambda^2$ is the coupling strength and $f(t) \propto |\phi( \mathbf{r}- \mathbf{v}_g t)|^2$ is    time-dependent envelope ($\mathbf{r}$ is the clock position), so that 
$  \int_{-\infty}^{\infty}  \delta \omega(t')  dt' = g \tau$.
 
 \begin{figure}[h]
\begin{center}
\includegraphics*[scale=0.5]{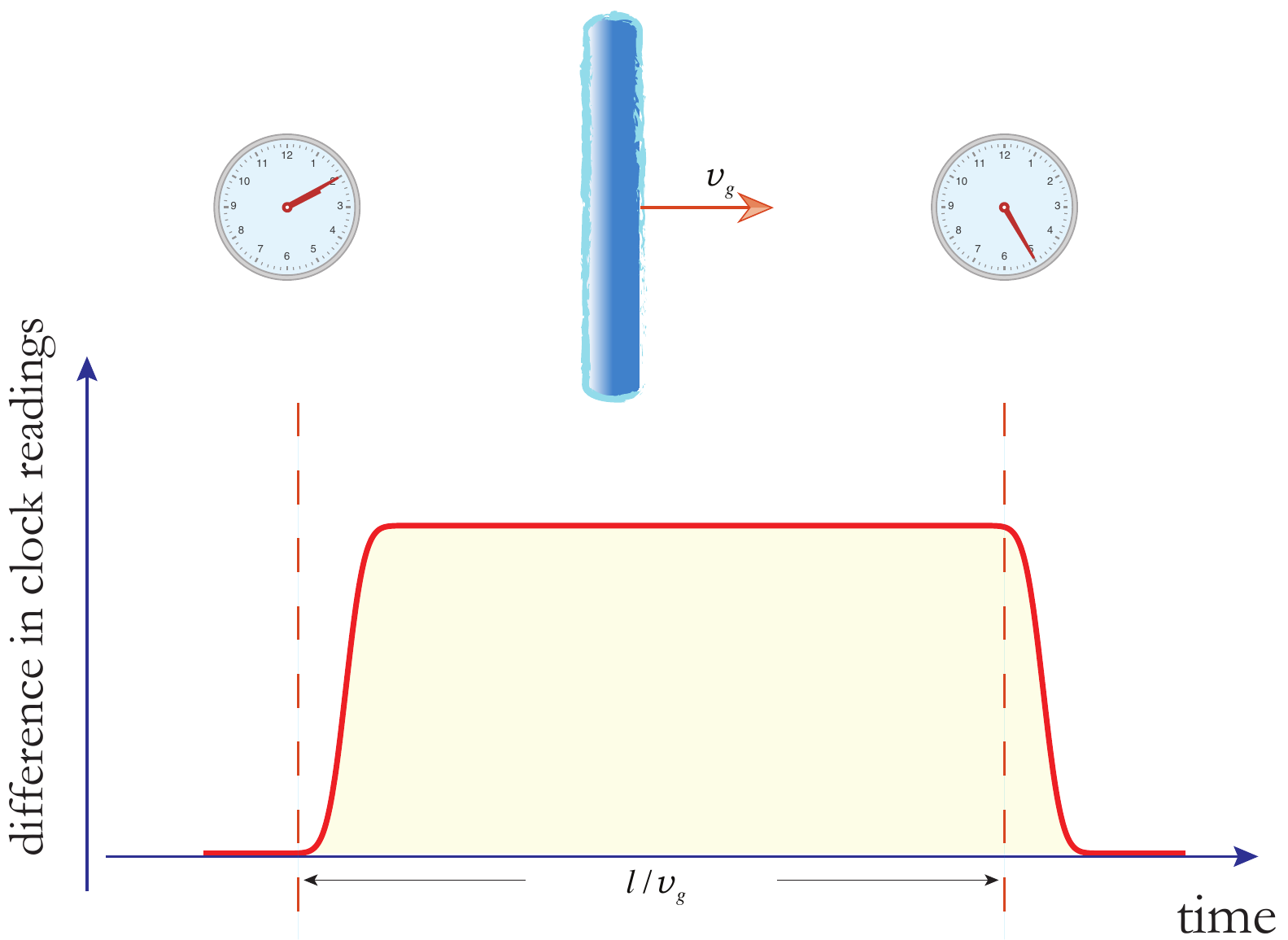}
\end{center}
\caption
{{\bf Concept of dark-matter search with atomic clocks.} By  monitoring  time discrepancy between   two spatially-separated clocks one could search for passage of topological defects,
such as domain wall pictured here.  
\label{Fig:Setup-wall}}
\end{figure}

Suppose we compare phases of two identical clocks separated by a distance $l$ (see Fig.\ref{Fig:Setup-wall})
that encounter a domain-wall-type TD. Because the TD propagates through the network with a speed $v_g$, 
the second clock would be affected by TD at a later time, with a time delay $l/v_g$. Formally,  the phase difference (or  apparent time discrepancy $\Delta t$) between the clocks reads  
 \[
 \Delta \varphi (t) = g  \int_{-\infty}^{t} (f(t'-l/v_g) - f(t')) dt'  \equiv \omega_0 \Delta t(t) \, .
\]
 By  monitoring correlated time difference $\Delta t(t)$ between  the two clocks, one could search for TDM.  Before the TD arrival at the first clock, the phase difference is zero, as the clocks are synchronized. As the TD passes the first clock, it 
picks
an additional phase  difference 
$ |\Delta \varphi|_{\max} = |g| d/ v_g$. $ \Delta \varphi (t)$   stays at that level while the TD travels between the two clocks. Finally, as the TD sweeps through the second clock, the phase difference vanishes. In this illustration we assumed that $d \ll l \ll L$. In the limit of $d \lesssim  l$ frequency (instead of time) comparison can be more accurate.

We may further relate the TD-induced frequency shift to the transient variation of fundamental constants. The instantaneous  clock frequency shift may be parameterized as 
\begin{equation}
 \frac{\delta \omega(t)}{ \omega_0} = \sum_X K_X \frac{ \delta X(t)}{X} \, ,  \label{Eq:variation}
\end{equation}
where $X$ runs over fundamental constants. 
The dimensionless sensitivity coefficients $K_X$ are known from atomic and nuclear structure calculations~\cite{FlaDzu09}.
It is important that different types of clocks exhibit sensitivity to different combination of fundamental constants,
with optical clocks being mostly sensitive to $\alpha$, and microwave clocks also to  nuclear couplings (see SI). 
 The energy density stored in the TD and various couplings enter  implicitly through time varying deviation, $\delta X(t)\propto |\phi( \mathbf{r} - \mathbf{v}_g t)|^2$, of the fundamental constant from its nominal value.
Then  the two clocks will be desynchronized by
\begin{eqnarray}
 |\Delta t |_{\max} & = & 
 \sum_X K_X  \int_{-\infty}^{\infty}  \frac{ \delta X(t)}{X} dt \sim \sum_X K_X \frac{A^2}{\Lambda_X^2} \tau \nonumber \\
  &\sim&   \sum_X K_X  \hbar c \frac{\rho_{\rm TDM} {\cal T}}{\Lambda_X^2} d^2  \, . \label{Eq:desync}
\end{eqnarray}
Here we used Eq.~(\ref{T}) and the fact that contributions to the Lagrangian  (\ref{interaction})  factorize into the SM and TDM parts. 
Notice that this result does not depend on a specific class of TDs.

In practice,  one needs to dissect the TD-induced desynchronization (\ref{Eq:desync})  from various noise sources  present in quantum devices and the link connecting the  two clocks.   We neglect link noise.
We assume that the TD thickness $d$ is much smaller than the distance between the clocks, as in Fig.~\ref{Fig:Setup-wall}. One would need to resolve the  ``hump'' in the presence of background noise.  Suppose we  compare the clock readings every $T$ seconds; then the total number of measurements of  non-zero phase difference is $N_m=l/(v_g T)$. 
For a terrestrial network with an arm length of $l \sim 10,000 \,  \mathrm{km}$, the TD sweep takes 30 seconds, so  one could make  30 measurements sampled every second.

Because the clocks are identical and statistically independent, the variance $\langle  \Delta \varphi (t) ^2  \rangle -  \langle  \Delta \varphi (t)  \rangle^2 = 2 R_\varphi(T)$, where $R_\varphi(T)$
is the phase   auto-covariance  function~\cite{Allan1966}.   It can be estimated  from the commonly reported Allan variance $\sigma_y (T)$, which characterizes fractional instability of the clock frequency~\cite{BarChiCut71}: $R_\varphi(T) \approx (\omega_0 T)^2 \,  \sigma_y^2 (T)$. Thereby  the uncertainty due to a single clock comparison is $\sqrt{2} (\omega_0 T) \,  \sigma_y (T)$. As we carry out $N_m=l/(v_g T)$ measurements, the statistical uncertainty is reduced further by $\sqrt{N_m}$.

The above argument leads to the signal-to-noise ratio
\begin{equation}
   S/N =  \frac{ c \hbar  \rho_\mathrm{TDM} \mathcal{T}  d^2 } 
    { T \sigma_y(T) \sqrt{  2 T v_g/ l }} 
     \sum_X K_X \Lambda_X^{-2} \, .
    \label{Eq:clockLimit}
\end{equation}
This ratio scales up  with the TD size $d$, the sensitivity coefficients  $K_X$, and the distance between the clocks. See SI for further discussion.

The TD detection confidence would improve with both increasing the number of network nodes and populating nodes with several clocks of different types. 
Clearly when the TD sweep is detected, all the clock pairs should exhibit time correlated desynchronization signature associated with the  sweep. Different clocks have distinct sensitivity to the variation of fundamental constants and this could help in disentangling various couplings in (\ref{interaction},\ref{Eq:variation}). Moreover, large number of clocks in a network will help determining the direction of the TD arrival, its velocity and spatial extent.

\begin{figure}[h]
\begin{center}
\includegraphics*[scale=0.5]{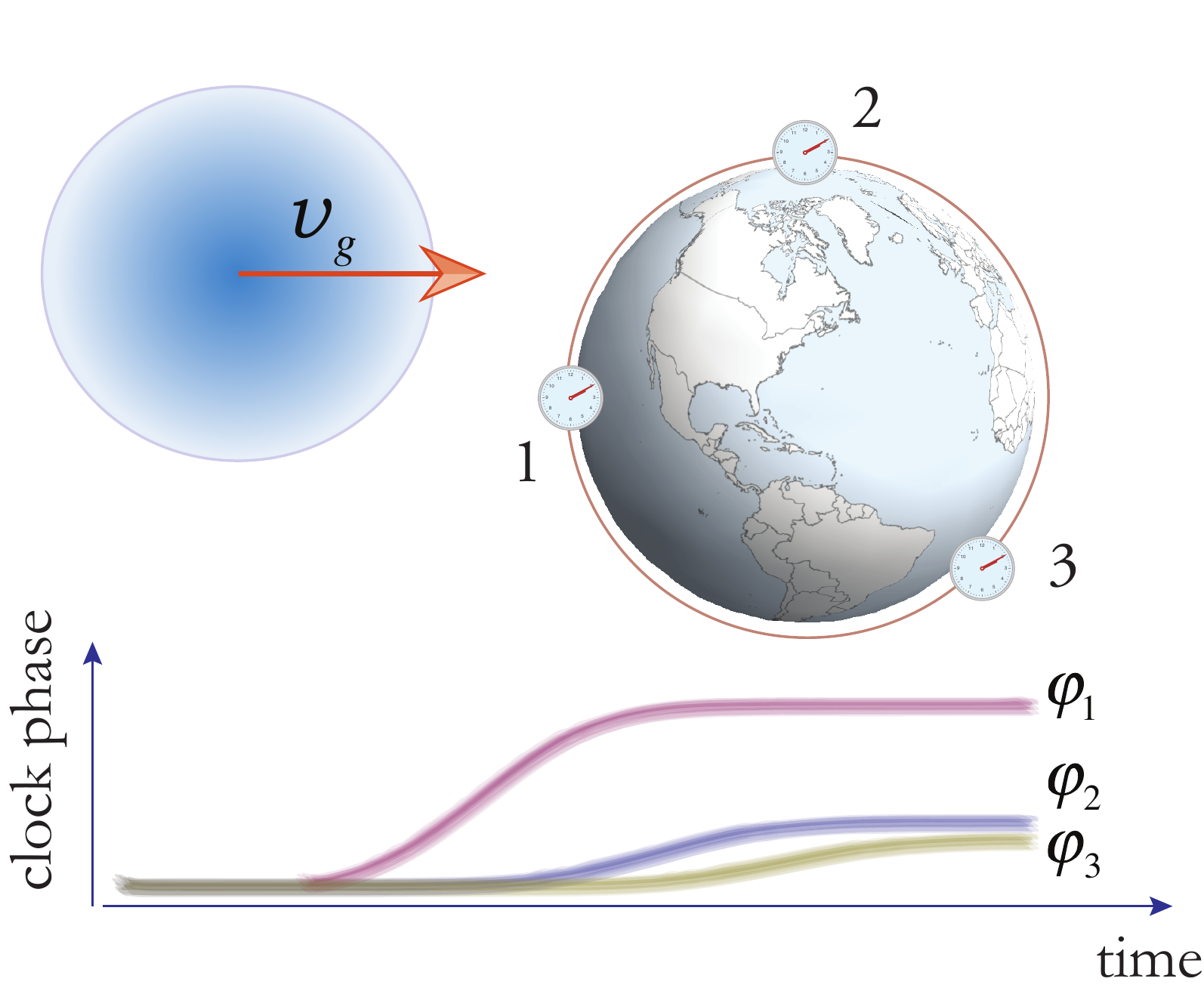}
\end{center}
\caption
{{\bf Effect of a monopole-type defect on atomic clocks.} Simulated response of an Earth-scale constellation of atomic clocks to a 0D Gaussian-profiled topological defect (monopole) of effective radius $0.75 R_\oplus$. The monopole center  is displaced from the collision axis by $0.2 R_\oplus$. The Earth center and the clocks lie in the collision plane. Polar angles of three clocks are $\pi/2, \pi, -\pi/4$ in the reference frame centered at the Earth center.
\label{Fig:Setup-globe}}
\end{figure}

The presented analysis can be generalized to the case of point-like TD (monopoles),
 which under gravitational force will behave identically to the
regular cold dark matter. 
We illustrate such a case in Fig. \ref{Fig:Setup-globe}. Here we assume that TD is an Earth-scale Gaussian-profile cloud sweeping  through a  clock network. Individual clocks  are perturbed at different times with different amplitudes, depending on the distance to the monopole center. This leads to a TD-induced phase accumulation,
\begin{eqnarray}
 \varphi_i (t) &=& g \int_{-\infty}^t  \exp\left\{ -(\mathbf{R}(t)-\mathbf{r}_i)^2/d^2 \right\} dt'  \\
&=& g_i    \int_{-\infty}^t  \exp\left\{ -(Z_0+ v_g t'- z_i)^2/d^2 \right\} dt'  \nonumber\, ,
\end{eqnarray}
where  $\mathbf{R}(t)=\{X_0, Y_0, Z_0+ v_g t \}$ and $\mathbf{r}_i=\{x_i,y_i,z_i\}$ are the TD center  and $i^\mathrm{th}$ clock positions and $d$ is the TD effective radius. Here we assumed 
that the TD propagates along the $z$-axis.
The coupling is rescaled depending on the clock position  $g_i  \equiv g \exp\left\{ -\rho_i^2/d^2 \right\} $, $\rho=( (X_0-x_i)^2 + (Y_0-y_i)^2)^{1/2}$ being the impact parameter.  
This translates into a differential phase accumulation between the clocks, similar to our ``wall'' example of Fig.~\ref{Fig:Setup-wall}, but with the step-on and step-off heights depending on the difference of clock impact parameters. 
Having several different-type clocks at each node of the network will maximize the discovery potential, increasing sensitivity to monopole and string-type objects, especially if their transverse size is much smaller than $R_\oplus$. In that case, direct comparison of several clocks within one node is needed. Implementing such a search with several clocks at a single node can be the first step towards a global TDM search effort. Detailed network optimization strategy for   TDM searches of varying transverse size $d$ and dimensionality $n$ are left for future investigations.

Several networks of atomic clocks are already operational. Perhaps the most well known are Rb and Cs microwave atomic clocks on-board satellites of the Global Positioning System (GPS)  and other satellite navigation systems.   We envision using the GPS constellation as a 50,000 km-arpeture dark matter detector,  with added capabilities due to extensive terrestrial network of atomic clocks on the GPS tracking stations.
 As  TDs sweep through the GPS constellation,  satellite clock readings are affected. Since accurate ephemeris satellite data are known, one could easily cross-correlate clock readings in the network.  For two diametrically-opposed satellites  the maximum time delay between clock perturbations would be $\sim 200 \, \mathrm{s}$, assuming the TD sweep with typical velocity of  300 km/s.  Different types of topological defects (e.g., domain walls  versus monopoles)  would yield distinct cross-correlation signatures. While the GPS is affected by a multitude of systematic effects, e.g., solar flares, temperature and clock frequency modulations as the satellites come in out of the Earth shadow,   none of conventional  effects would propagate with  300 km/s through the network.

 Dark matter search can also be implemented with the state-of-the art laboratory clocks \cite{ChoHumKoe10,BloNicWil14}, 
utilizing vast network of atomic clocks at national standards laboratories used for evaluating the TAI timescale~\cite{Lev99}.  Moreover, several elements of high-quality optical links for clock comparisons have been already demonstrated in Europe,
with 920 km link connecting two laboratories in Germany~\cite{Predehl2012}.
 In addition, a Cs fountain clock and an H-maser are planned to be installed on the international space station in the near future, providing high-quality time and frequency link to several 
metrology l laboratories around the globe~\cite{ACES}. 

\begin{figure}[h]
\begin{center}
\includegraphics*[scale=0.62]{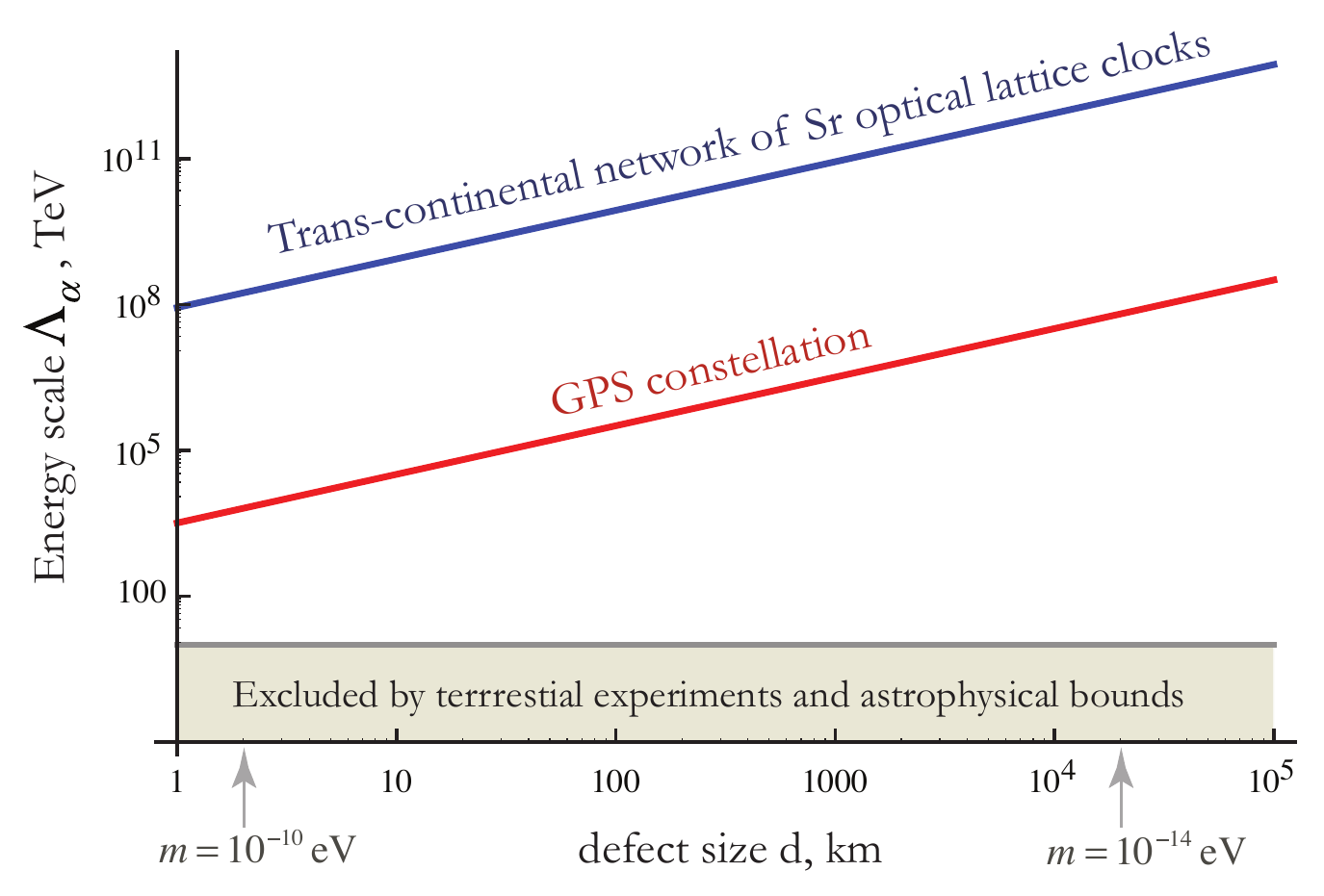}
\end{center}
\caption
{ {\bf Projected constraints on dark-matter coupling.} Terrestrial and space networks of atomic clocks can impose powerful constraints on characteristic energy scales of dark-matter interaction with baryonic matter~(\protect\ref{interaction}). Here we show bounds on $\Lambda_\alpha$ that may be derived from a terrestrial network of optical lattice clocks and GPS clocks. The horizontal axis is the topological defect size in km and also includes two characteristic TD field rest mass scale values.
\label{Fig:Sensitivity}}
\end{figure}

As an illustration of sensitivity to energy scales $\Lambda_X$ of TDM-SM coupling (\ref{interaction}), we consider a terrestrial network ($l\sim10,000$ km) of Sr optical lattice clocks which are sensitive to the variation of $\alpha$ with $K_\alpha = 6 \times 10^{-2}$. For these clocks one may anticipate reaching $\sigma_y(1 \, \mathrm{s}) \sim 10^{-18}$ at  $T=1\,\mathrm{s}$ measurement intervals. Requiring $S/N \sim 1$ in Eq.(\ref{Eq:clockLimit}), substituting fiducial values for $\rho_{\rm TDM}$ and $v_g$, and choosing ${\cal T} \sim 1$ yr, we draw sensitivity curve to the energy scale $\Lambda_\alpha$ as a function of the defect size in Fig.~\ref{Fig:Sensitivity}.  Here we also show sensitivity of GPS constellation ($l\sim50,000$ km, $T=30 \, \mathrm{s}$, $\sigma_y(30 \, \mathrm{s})\sim 10^{-11}$) assuming that the TDM-SM coupling is dominated by the transient variation of $\alpha$ ($K_\alpha =2$). Limits derived from both Sr and GPS networks would greatly exceed the  $\Lambda <10 \, \mathrm{TeV}$ region excluded  by direct laboratory and astrophysical constraints, such as from fifth-force and the violation of the equivalence principle searches~\cite{Olive:2007aj}.

{\it Acknowledgment:}
We would like to thank N. Fortson, P. Graham, J. Hall, M. Murphy, J. Sherman, J. Weinstein, and I. Yavin for discussions.  This work was supported by the US National Science Foundation. \\

\section{Supplementary Information}

The following is more in-depth exposition of relevant details and background material 
that could not be included in the main 
text due to the length restriction. 

\subsection{ Dark Matter and Dark Energy }

The most precise measurements of the cosmological parameters by the Planck satellite \cite{Ade:2013zuv} determine the 
global abundance of Dark Matter (DM) and Dark Energy (DE) parameters (see {\em e.g.} Ref. \cite{RPP-2012}):
\begin{equation}
\rho_c  = 1.05\times 10^{-5} h^2 ~{\rm GeV~ cm^{-3}}; ~~\rho_{\rm DM} = 0.22 \times \rho_c ; ~~\rho_{\rm DE} = 0.73 \times \rho_c
\end{equation}
We use $\hbar=c=1$ throughout, and $h\simeq 0.7$ is the  Hubble expansion rates in units of 
$100 \, {\rm km \,s^{-1}\,Mpc^{-1}}$. It is remarkable that only small remainder of $\sim5\%$
from the total energy balance is contributed by the ordinary matter, such as atoms and radiation,
which is perfectly described by the Standard Model (SM) of particles and fields. 
The principal difference between DM and DE is in their clustering properties: the DM has no detectable pressure and is 
responsible for the formation of the cosmological structure, while DE has negative pressure that causes the
accelerated expansion of the Universe. 

The DM is 
believed to be responsible for the formation of cosmic structures, and in particular of the Milky Way galaxy. 
The energy density of galactic DM, at the approximate location of the Solar System is measured to be \cite{RPP-2012} 
around 
\be
\rho_{\rm DM}^{\rm gal} \simeq 0.3~{\rm GeV~ cm^{-3}}
\ee
with a factor of $\sim 2$ uncertainty. 
It must be noted that the direct gravitational constraints on the energy density of the 
dark matter inside the Solar System are less restrictive, giving 
the limit on DM density within 1 A.U. \cite{Pitjev:2013sfa} of
\be
\rho_{\rm DM}^{\rm Solar} < 10^5~{\rm GeV~ cm^{-3}}.
\ee

From the observation of clustering of matter on different astronomical scales, 
it is known that DM consists of non-relativistic objects, but neither typical masses nor linear dimensions of DM objects 
are known. This is the direct consequence of the universality of
gravitational interactions, which does not introduce any sensitivity to the types of gravitating objects.
If dark matter consists of some self-similar constituents of mass $m_{\rm DM}$ and 
number density $n_{\rm DM}$, the above expressions for the DM energy density only fix the 
product of the two: $\rho_{\rm DM} = m_{\rm DM} n_{\rm DM}$. Even if we {\rm assume} that DM 
consists of elementary particles, it leaves an enormous freedom in the possible 
choice of $m_{\rm DM}$, that extends across fifty orders of magnitude. 
\begin{equation} 
10^{-22}~{\rm eV} < m_{\rm DM} < 10^{28} ~{\rm eV} \, .
\end{equation} 
The lower limit in this relation comes from the inverse halo size of smallest galaxies, and 
the upper limit is set by the requirements that these particles not form black holes. 
An extensive experimental program aimed to detect DM particles with masses 
commensurate to the masses of SM elementary particles, $O(1-10^3)$ GeV, via the elastic scattering 
of DM on nuclei, has not resulted in a positive detection. The most 
precise experiment to date, LUX, have recently reported negative results, constraining the scattering cross section 
per nucleon to below $10^{-45}\,{\rm cm}^2$ for the optimal mass choice for  a DM particle \cite{Akerib:2013tjd}. 
The absence of any credible signals from particle DM candidates calls for {\em broadening } the scope of 
DM searches.  

Once the assumption of DM being composed from elementary particles is 
relaxed, the  possible freedom in $m_{\rm DM}$ parameter can be extended  above 
$10^{28} ~{\rm eV}$, if one considers DM composed of the 
spatially extended objects. They can consist from agglomeration of 
dark particles brought together by some dark force, and protected from annihilation by  
additional symmetries. This would be  similar to baryons and leptons of the SM sector
combining to the bigger objects such as grains of dust, planets, stars etc. 
It may also happen that DM objects are formed by stable field configurations
of some light fields in the form of topological defects, such as monopoles, strings and domain walls. 
We will call the dark matter composed from such objects as ``topological dark matter'', or TDM, 
noting that most of our considerations apply also to other forms of DM objects  with large spatial extent.

The flip-side of gravity being insensitive to the masses of DM objects, is a robust expectation
for the velocity of DM {\em regardless} its exact nature.  It is widely believed that in the galactic rest frame, the 
distribution of DM objects over  velocities is quasi-Maxwellian, with the velocity dispersion 
$v\simeq 270\, {\rm km \,s^{-1}}\simeq 10^{-3}$, and the sharp cutoff above the galactic escape velocity of 
$v_{\rm esc} \simeq 650\, {\rm km \,s^{-1}}$ (see, {\rm e.g.} Ref. \cite{Savage:2006qr}). From the point of an observer bound to the 
Solar System, this distribution is modified by an addition of the sun's motion relative to the galactic 
center, that is well known to be $230\, {\rm km \,s^{-1}}$. If in addition to the galactic dark matter, 
there is a population of DM objects bound to the 
Solar System, then it is natural to expect that their velocities will be one order of magnitude smaller, that is comparable to the 
planetary velocities. 

The existence of these priors on DM velocity distribution is very important, as it allows to calculate the statistical 
expectation for the frequency and the geometry of the DM encounter with any detector, once $m_{\rm DM}$ and the linear 
dimensions of DM objects are specified.

\subsection{DM interaction portals }

In order to study the DM-SM interaction, one has to specify how two sectors interact. 
We use a now standard approach of so-called portals, when the gauge invariant operators of the 
SM fields are coupled to the operators that contain fields from the dark sector. 
This phenomenological approach is widely used in particle physics for searches of DM and dark forces, see
{\em e.g.} recent review \cite{Essig:2013lka}. In this section, we list leading candidates for the portal interactions, 
and discuss the constraints on their strength. 

{\em Types of portals}\\
In enumerating possible interactions, we shall assume that the DM objects are built from the scalar fields. 
The generalization to other types of fields is also possible. We call the fields from the DM sector as $a$, 
while the SM fermions are called $\psi$. The portal interactions are individual terms ${\cal L}_i$, and their
sum defines the total DM-SM interaction Lagrangian, ${\cal L} = \sum_i {\cal L}_i$. Listing only a few leading terms, we have 
\begin{eqnarray}
{\cal L}_1 = \frac{\partial_\mu a}{\Lambda} \sum_{\rm SM~particles } c_\psi \bar\psi \gamma_\mu\gamma_5 \psi  ~~~~&&{\rm axionic~portal}
\\
{\cal L}_2 = \frac{ a}{\Lambda} \sum_{\rm SM~particles } c^{(s)}_\psi m_\psi \bar\psi  \psi  ~~~~&&{\rm scalar~portal}
\\
{\cal L}_3 = \frac{ a^2}{\Lambda^2} \sum_{\rm SM~particles } 
 c^{(2s)}_\psi m_\psi \bar\psi  \psi  ~~~~&&{\rm quadratic ~ scalar~portal}
\\
{\cal L}_4 = \frac{ia^* \partial_\mu a }{\Lambda^2 } \sum_{\rm SM~particles } g_\psi \bar\psi \gamma_\mu \psi
~~~~&&{\rm current-current~portal}
\end{eqnarray}
Here $\Lambda$ is the energy scales, and
$c_i$ are individual coefficients that can take on different values depending on type of $\psi$. Each of these interactions 
implies new physics at the scale $\Lambda$, but since we are taking it above the weak scale, 
we are not required to provide an explicit UV-completion. This classification can be generalized to include the 
SM gauge bosons, via $m_\psi \bar\psi  \psi\to F_{\mu\nu}^2$ substitutions. 

For the DM search with atomic clocks the interactions ${\cal L}_2$ and ${\cal L}_3$ 
are the most important, as they provide a shift of the SM particle masses and coupling constants inside a
DM object. We note that other portals are also of interest but for the different types of probes: for example 
both ${\cal L}_1$ and ${\cal L}_4$ induce a magnetic-type shift inside a defect and can be searched for with the 
magnetometry techniques. 

{\em Direct laboratory and astrophysical constraints} \\
Since the DM fields $a$ involved in our considerations are light, 
the scalar portals, ${\cal L}_2$, ${\cal L}_3$, and their $F_{\mu\nu}^2$-proportional modifications 
can be constrained in the direct experiments, and via the modifications of 
astrophysical processes. To indicate the scale of these constraints, we shall take the values of 
all coefficients $c_i\simeq 1$, and set the constraints on $\Lambda$. 

The constraints on ${\cal L}_2$ are quite strong, and given by
\begin{eqnarray}
{\cal L}_2:~~ \Lambda > 10^{12} {\rm GeV, ~astrophysics}; ~~\Lambda > 10^{21} {\rm GeV, ~gravity ~tests.}
\end{eqnarray} 
They come from the considerations of the energy loss to $a$-quanta in stellar processes \cite{Raffelt:1999tx},
and from the very precise tests of the gravitational interactions \cite{Bertotti:2003rm}. 
Indeed, the scalar interaction ${\cal L}_2$ induces $1/r$-type attractive potential between the SM particles, and has different 
post-Newtonian corrections compared to general relativity. In contrast to ${\cal L}_2$, ${\cal L}_3$ can provide 
only the $1/r^3$ attractive potential, as only the exchange by a pair of $a$'s
is allowed by the form of this portal. Consequently, all 
constraints are considerably milder \cite{Olive:2007aj},
\begin{eqnarray}
{\cal L}_3:~~ \Lambda > 10^3-10^{4} {\rm GeV, ~astrophysics~and ~gravity ~tests.}
\end{eqnarray} 
On account of these mild limits, it is the quadratic scalar portal that is the most suitable for 
searches with the use of atomic clocks. More specifically, the supernova energy loss bounds limit 
$\Lambda_{e,\gamma} > 3$~TeV and $\Lambda_{p} > 15$~TeV, while tests of the gravitational 
force can be used to limit  $\Lambda_p > 2$~TeV \cite{Olive:2007aj}. We are not going to differentiate between 
these bounds, and will assume the that the existing reference sensitivity for $\Lambda_X$ is 10~TeV.

{\em Questions of technical naturalness} \\
It is well-known that portals with non-derivative interactions face a problem 
of fine-tuning at the level of radiative corrections. In particular, interactions encoded in 
${\cal L}_2$ and ${\cal L}_3$ at the quantum level may induce corrections to the mass parameter
of the $a$-field, that is sensitive to the ultraviolet cutoff, $\Delta m_a^2 \sim m_\psi^2 \Lambda_{\rm UV}^2/\Lambda^2$.
The problem arises when one requires physical mass of the $a$ field to remain light, while 
pushing $\Lambda_{\rm UV}$ to the scale of $\Lambda$. 
This is a well-known theoretical problem, and a common drawback to almost {\em any} model that involve 
light scalar fields. While we do not attempt to solve it here, we notice that for the idea of the TD dark matter
the problem is less severe than in any models that predicts a continuous change of masses and coupling constants in 
time. Indeed, for the latter to happen the mass parameter would have to be maintained at the level comparable to the 
Hubble expansion rate, $10^{-33}$ eV, or below. This is almost impossible to reconcile with any realistic choice of the cutoff
\cite{Banks:2001qc}. In comparison, typical scales of $m_a \sim 10^{-13}$ eV  used in this work lead to a much 
less restrictive fine-tuning problem. Thus, from field theoretical point of view, the transient change of the coupling constants
is more "natural" than a constant slow drift.

\subsection{Topological defects contributing to DM and DE} 

The cosmological evolution of the early Universe involves expansion and cooling of the 
primordial matter. It is tempting to think by analogy with condensed matter systems, that 
during that process a number of the phases transitions have occurred. One transition, from 
the unbroken Higgs phase at high temperature to the broken familiar phase at low temperature 
have happened at the electroweak epoch, $T\sim100$ GeV; another transition, from a deconfined to a confined phase 
occurred at temperatures commensurate with the energy scale of strong dynamics, $T \sim 200$ MeV. It is then 
very natural to think that if the full theory contains additional field-theoretical structures, such as Grand Unified Theory (GUT), 
Peccei-Quinn sector responsible for axions, or more generically some ``dark'' sector giving rise to DM, 
there can be additional phase transitions beyond the QCD and electroweak transitions. 

The importance for cosmology of phase transitions and the resulting topological defects akin to objects 
such as vortices, dislocations and domain walls, very familiar to physicists from condensed matter examples, 
was understood early on. A wealth of cosmology literature pursued the idea of structure formation seeded by the 
cosmic strings, remnants of some early cosmological phase transitions \cite{Kibble:1980mv,Vachaspati,Vilenkin:1984ib,Zurek:1985qw,Hindmarsh:1994re,Durrer:2001cg}. 
However, after the precise measurements of the 
CMB anisotropies, it became clear that the spectrum of density perturbations is well described by inflationary models, 
and cosmic strings as well as other types of defects could have played only a secondary role in generation of density perturbations. 
Nevertheless, the contribution of topological defects in the dark sector to the DM and DE can be significant, 
and several ideas for both microscopic  and macroscopic realization of such possibility has been suggested recently 
\cite{Baek:2013dwa,Evslin:2013mga}. A question of significant interest to us is whether one can contemplate
objects of significant spatial extent, amenable to search strategies proposed in the main body of our 
manuscript. Below we describe the salient features of the TDs and present some simple estimates for their abundance.

{\em Domain walls, strings and monopoles} \\
Consider the scalar field $\phi$, either real, $\phi=\phi^*$, or complex,  that has a 
simple self-interaction potential
\be
\label{V}
V(\psi) = \lambda( |\phi|^2- A^2)^2. 
\ee
If the field is real, there exist two energetically equivalent vacua, $\langle \phi \rangle = \pm A$, and if two domains are 
present, then a domain wall with $\phi(x) = A\tanh(mx)$  arises at the interpolation region between them. The thickness 
of domain wall is given by the inverse of Higgs-like mass parameter, $d \simeq m^{-1 }= (\sqrt{\lambda} A)^{-1}$, 
characterizing the stiffness of the potential near the minimum. Notice that a macroscopic size of the domain wall can be achieved either with small 
$\lambda$ or small $A$, or both. If the form of $V(\phi)$ is more complicated and allows for more than two 
degenerate minima, a network of domain walls separating vacua with different 
$\langle \phi \rangle$ is allowed to form.  

The same potential gives rise to a global string solution if $\phi$ is a complex field,
as there is a continuum of vacuum solutions, different only by a phase. 
 The model of the complex scalar field can also be gauged, in which case 
the string solution will also support a non-vanishing "magnetic" flux of the gauge field. The transverse dimension
of the string are set by the parameters $\sqrt{\lambda} A$ and $gA$, where $g$ is the gauge coupling. 
A non-Abelian version of (\ref{V}), with {\em e.g.} $V(\phi) =  \lambda( \phi^ba^b- A^2)^2$, with adjoint $a^b$ 
charged under the $SU(N)$ gauge group is known to develop the monopole solutions \cite{Polyakov:1974ek,'tHooft:1974qc}. 
In general, if the vacuum manifold ${\cal M}$ in the 
field space $a$ allows for the disconnected regions, then the stable domain walls can form.
 If ${\cal M}$ allows for the existence of non-contractable loops, 
then strings can form, and existence of non-contractable spheres will lead to monopole solutions (see {\em e.g.} Ref. \cite{Durrer:2001cg}). 
It is important to note that in many theories the higher-dimensional defects can decay to the lower-dimensional ones, forming the 
sequence of transitions: walls$\to$strings$\to$monopoles.

If a frustrated (self-similar) network of the domain walls, strings or monopoles is created in the course of a phase transition, 
then the energy density of the network would scale according to 
\be
\label{ansatz} 
\rho_{\rm network}(t) \propto \frac{A^2 d^{n+1}}{[L(t)]^{n+3}},
\ee
Here $n=0,1,2$ for the monopoles, strings and domain walls, $d$ is the thickness of a TD, and $L(t)$ is an approximate
distance between neighboring defects. In the regime when $L(t)$ evolves together with the cosmological scale factor $R(t)$, 
networks can be assigned the effective equation-of-state parameter, $w_{\rm network} = p/\rho= - n/3$. One can see then 
that the gas of monopoles would automatically mimic pressureless fluid, identical in its clustering properties to any cold dark matter fluid. 
Therefore, monopoles are ideal cold dark matter candidates, if their initial abundance matches $\rho_{\rm DM}$. 

Networks of strings and 
domain walls would have an equation of state $w=-1/3$ and $-2/3$. The latter was considered to be one of the candidates for DE 
\cite{Battye:1999eq}, before the equation of state parameter was measured experimentally to be close to $w\simeq -1$.
Current measurements of $w$ allow domain walls as only a subdominant source of the negative pressure compared to the 
cosmological constant, but domain walls are still allowed to contribute to up to $O(30\%)$ to $\rho_{\rm DE}$. 
The scaling ansatz (\ref{ansatz}) may be too simplistic, as it neglects processes of walls and strings collisions and inter-connections. 
For example, string collision can lead to the formation of string loops, which generically result in steeper  fall-off with $R(t)$. 
In this case, effective $w$ of string and wall networks can be closer to $w\simeq 0$, and thus similar to the 
non-relativistic fluid, allowing to consider these objects as a candidate for $\rho_{\rm DM}$. We also notice that 
a much smaller energy density inside walls and strings in our scenario, compared to the defects built from heavy GUT fields, 
would allow for a longevity of closed string loops, protecting them from rapid decays with the emission of the SM particles. 

{\em $Q$-balls and other extended objects}\\
While strings and domain walls would likely require topological protection of their stability, there are more 
ways other than topology for achieving stable extended objects similar to monopoles in their clustering properties. 
For example, one can construct models with 
DM built from a new Dirac field $\psi$, with the number of particles and antiparticles 
being different, $N_{\psi}-N_{\bar\psi}>0$. If there is an additional attractive force in the $\psi$-sector, 
the space regions with condensates of many $\psi$ particles will form, and the extent of these regions
can be many orders of magnitude larger than the Compton wave length of $\psi$. 

A well-studied idea is the non-topological condensate of the scalar field $\phi$ charged under the new $U(1)_X$ force. 
If there is a global $X$-charge asymmetry of the Universe, it is energetically preferable to 
form new type of non-topological configurations of $\phi$ field  that cary a large $X$-charge, $Q\gg 1$. 
Such defects are called Q-balls \cite{Lee:1991ax,Coleman:1985ki}. Unlike solitons, Q-balls have  time-oscillating fields inside, and 
do not have to be "identical", {\em i.e.} admitting the distribution over possible $Q$-charge and the size. 
It is also well known that they can 
be a good candidate for the DM \cite{Kusenko:2001vu}. As in the literature on TD, it is usually assumed that the 
Q-balls are built from relatively heavy field (e.g. scalar supersymmetric partners of the SM particles \cite{Kusenko:1997zq}),
but this does not have to be the case. Also importantly, the physical
size  of the Q-balls is sensitive to the form of the potential $V(|\phi|)$. Potentials that admit a 
near flat direction are known to develop Q-ball solutions that have {\em both} large spatial extent {\em and} 
large amplitude of the field inside, $A,d\propto Q^{1/4}$ \cite{Dvali:1997qv}. Thus macroscopic size  Q-balls
that form DM can be an ideal candidate for the search with the use of the atomic clock networks. 

{\em Example of cosmic string with an increased $\alpha_{EM}$ in its core.}\\
In this subsection we present an important case of a TD interacting with the SM, that does not 
require any additional UV completion and  is by itself a self-contained example. 
This example is based on  the so-called "dark photon" \cite{Holdom:1985ag}  theory, 
which represents an abelian group $U(1)_{\rm dark}$, with a scalar field $H$ that forms $\langle H\rangle = v \neq 0$ condensate,
giving dark photon a mass. It is well known that such a theory admits 
TD in form of a string-type defect, in the middle of which the expectation 
value of the Higgs field vanishes and the gauge symmetry is restored. 
That way, $m_V$, the mass of the dark photon $V_\mu$ varies across the defect 
as a function of $r=(x^2+y^2)^{1/2}$, 
exactly vanishing in the middle at $x=y=0$.
The coupling between dark photon and the electromagnetic field is introduced via a small 
mixing angle $\kappa$. 
Thus the whole Lagrangian of this extension of the SM is given by
\begin{eqnarray}
{\cal L} = -\frac14 F_{\mu\nu}^2  -\frac14 V_{\mu\nu}^2-\frac{\kappa}{2} F_{\mu\nu} V_{\mu\nu} +|D_\mu H|^2 - U(|H|^2)
\nonumber\\
\rightarrow   -\frac14 F_{\mu\nu}^2  -\frac14 V_{\mu\nu}^2-\frac{\kappa}{2} F_{\mu\nu} V_{\mu\nu} +\frac{1}{2} [m_V(x,y)]^2V_\mu^2.... 
\end{eqnarray}
In this formula, $V_\mu$ is the dark photon field, $F_{\mu\nu}$ and $V_{\mu\nu}$ are the fields strength for the electromagnetic and 
dark gauge groups,  $D_\mu = \partial_\mu + ig'V_\mu$ is the  covariant derivative, 
and $U(|H|)$ is the dark Higgs self-potential. In the second line of this formula, a reduction to the broken phase is 
performed, and $m_V(x,y)= g'\langle H(x,y) \rangle$ is the space-varying mass of the dark photon. 
The string thickness is set by the mass of the Higgs boson, $d^{-1} \sim m_H$.  

The effective strength of the electromagnetic force will vary as a function of position relative to the center of the string. 
It is easy to see that on top of the usual $1/r$ Coulomb interaction  
due to the exchange by the EM field, there comes a Yukawa-type correction from the photon-dark-photon mixing.
Considering for concreteness the interaction of two point charges $q_1$ and $q_2$  separated by a typical atomic distance distance $a$ ($a\sim 
\frac{\hbar^2}{m_ee^2} \ll d$), in the first order in mixing angle $\kappa$, one has
\begin{equation}
V({ a}) \simeq \frac{q_1 q_2}{a}\left(1 + \kappa^2 \exp\left[-a m_V \right]\right)
\end{equation}
Outside the string one can have $m_V a \gg 1$, while inside $m_V \simeq 0$ and  $\exp[-a m_V] \simeq 1$, leading 
to the effective difference of the electromagnetic coupling constant $\alpha$, 
\begin{equation}
\frac{\alpha_{\rm inside}(x=y=0)}{\alpha_{\rm outside}(r >d)} \simeq 1+\kappa^2.
\end{equation}

The model of dark photons has been a subject of intense theoretical investigations and experimental searches, and the 
limits on its parameter space ($m_V$, $\kappa$) are rather well-known \cite{Essig:2013lka}: while for $m_Vc^2 < 10$ keV 
there are strong bounds from stellar energy loss, for heavier $m_V$ sizeable mixing angles can be allowed, 
and in particular, $\Delta \alpha/\alpha = \kappa^2 > 10^{-10}$ for $m_V > 10$~MeV is not ruled out. 
Therefore, if in the broken phase of the dark gauge symmetry $m_H \ll m_V$ (which in turns imply the 
value of the Higgs self-coupling $\lambda$ be much smaller than the gauge coupling, $\lambda \ll (g')^2$),
one can have a macroscopic size TD, affecting microscopic corrections to the Coulomb law,
 resulting in a relatively large, $O(10^{-10}$) or more,  shift in the fine structure constant. 
Thus, crossing of the dark photon string TD, that has a slightly larger value of $\alpha$ inside its core, 
will lead to a transient shift of atomic frequencies. 

{\em Remarks on cosmological abundance of TDM} \\
There are two general mechanisms for generating the non-zero abundance of solitonic objects, of 
topological nature or not. The first one is based on the idea of fragmentation into different domains 
followed the phase transition, $\langle \phi \rangle =0 \to \langle \phi \rangle =A\exp\{i\theta\}$, 
generically referred to as Kibble mechanism. This picture assumes random distribution over $\theta$,
when averaged over the larger (super)-horizon size space-time patches at the time of the phase transition. 
The alternative mechanism uses the idea of fusion of small non-topological solitons 
into bigger objects (see {\em e.g.} \cite{Griest:1989bq}), and is often applied for estimation of the Q-ball abundances.
Notice in the case of Q-balls, there is an additional conserved quantity, the charge asymmetry under $U(1)_X$, 
that can be always adjusted in order to create a desirable abundance. 

In the concluding paragraphs we will estimate the abundance of monopoles assuming the following simplified framework: 
the time of the TD formation coincides with the cosmological epoch when the size of the Hubble horizon becomes 
comparable to the size of the defect, $H(T_{form}) d\sim 1$. Here $T_{form}$ is the temperature of the 
cosmological fluid at the time of the defect formation. 
At that point approximately one monopole per Hubble volume is 
formed, and the subsequent evolution of the energy density is given by the usual scaling for non-relativistic particles, 
$\rho_{\rm TDM} \propto (R(t))^{-3}$. The CMB anisotropies give plenty of evidence that the DM existed prior to the CMB decoupling, 
and at least from the matter-radiation equality. Thus, $T_{form}$ has to be less than the matter-radiation equality temperature, 
$T_{eq} \simeq 1$eV. In order to be a significant component of the DM, monopoles would have to satisfy 
the following approximate relation, 
\begin{equation}
\frac{A^2d^{-2}}{\rho_{rad}(T_{form})}\times \frac{T_{form}}{T_{eq}} \sim 1.
\end{equation}  
 Moreover, the energy density of radiation at the time of the formation of monopoles can be expressed 
via the Hubble rate, and ultimately via $d$, 
\be
\frac{8\pi G}{3}\times  \rho_{rad}(T_{form}) =  H(T_{form})^2 \sim d^{-2},
\ee
leaving us with 
\be
\label{final} 
\frac{8\pi A^2 }{3 M_p^2}\times \frac{T_{form}}{T_{eq}} \sim 1,
\ee 
where the Newton constant $G$ was substituted for the inverse Planck mass squared. Because the Planck mass $M_{p}$ 
is very large, this condition can be satisfied only if $A$ or $T_{form}$ are large as well. 
It turns out that it is difficult to satisfy this condition for the most minimal model of the defect with 
one scale for the amplitude of the $a$-field inside, and the fiducial choice of parameters, $d\sim R_{Earth}$, 
and time between TD encounters ${\cal T}\sim$ 1 yr. The problem arises because $T_{form}$ comes out 
too low. This problem can be circumvented in a model where a monopole with amplitude $A$ and size $d$ 
contains a hard core of a small radius $d_{core}$, with a large 
field inside, $A_{core}$, $A_{core}\gg A;~d_{core} \ll d$. In that case large $T_{form}$ can be achieved due to the smallness of $d_{core}$, and the equation on the abundance (\ref{final}) can be easily satisfied. In other words, models of TD 
have enough flexibility to arrange for a realistic cosmological scenario for TD as the leading contributor to DM. 
We defer exact details of this model for future investigations, and conclude by simply stating that indeed cosmological 
models with realistic TDM  abundance and parameters amenable to searches with atomic clocks 
 can be constructed. 

\section{ Sensitivity of atomic clocks to TDM}
In this section we compare sensitivity of various atomic clocks to  TDM. 
We start with reproducing the signal-to-noise ratio, Eq.~(6) of the main text,
\begin{equation}
   S/N =  \frac{ c \hbar  \rho_\mathrm{TDM} \mathcal{T}  d^2 } 
    { T \sigma_y(T) \sqrt{  2 T v_g/ l }} 
     \sum_X K_X \Lambda_X^{-2} \, .
    \label{Eq:clockLimit}
\end{equation}
(We restore $\hbar$ and $c$ in this section to be more consistent 
with atomic physics notation.)

An important ingredient in this formula are the sensitivity coefficients $K_X$~\cite{FlaDzu09},
which we review in the following. One could parameterize the variation of the clock frequency as
\begin{eqnarray}
 &\frac{\delta(\omega_0/U)}{ \omega_0/U} =& \frac{\delta V}{V} \, ,\nonumber \\
 &V =&  \alpha^{K_\alpha} \left( \frac{ m_q}{\Lambda_\mathrm{QCD}} \right)^{K_q}  \left(\frac{m_e}{m_p}\right)^{K_{me/p}} \, . \label{Eq:DetailedCoeff}
\end{eqnarray}
Here $U$ is the unit of angular frequency, $m_q$ is a quark mass, $\Lambda_\mathrm{QCD}$ is the quantum chromodynamics mass scale, and $m_e/m_p$ is the electron to proton mass ratio. The powers of various factors are the sensitivity coefficients $K_X$ that enter Eqs.~(4-6) of the main text. The sensitivity coefficients enter Eq.~(\ref{Eq:clockLimit}) in a particular combination
\be
 S = \sum_X K_X \Lambda_X^{-2} \,. \label{Eq:sumS}
\ee

In general, one distinguishes between two broad classes of atomic clocks: microwave and optical clocks.
Microwave clocks, such as H, Rb and Cs, operate on hyperfine transitions,  the frequency of such transitions being determined by the coupling of atomic electrons to nuclear magnetic moments and thereby these depend both on $\Lambda_\mathrm{QCD}$ and $\alpha$. For microwave clocks, $K_\alpha$ ranges from $2$ in $^1$H to 4.28 in heavy $^{199}\mathrm{Hg}^+$~\cite{FlaDzu09}; $K_\alpha$  grows with nuclear charge due to increasing  relativistic effects. Nuclear-structure-dependent coefficient $K_q$ exhibits non-monotonic behavior~\cite{FlaTed06}: $K_q=-0.09$ for $^1$H, $0.11$ for $^{133}\mathrm{Cs}$, and $-0.12$ for $^{199}\mathrm{Hg}^+$. $K_{me/p}=1$ for hyperfine transitions. Thus for the microwave clocks the sum $S$, Eq.~(\ref{Eq:sumS}),  contains three terms. For example, for  hydrogen masers,
\be
 S(^1H) = \frac{2}{\Lambda_\alpha^2} - \frac{0.09}{\Lambda_q^2} +   \frac{1}{\Lambda_{me/p}^2} \,. \label{Eq:H-sensitivities}
\ee

As to the optical clocks, here the dependence on fundamental constants is simplified, as these clocks utilize electronic transitions and their clock frequency depends only on $\alpha$ ($K_q=K_{me/p}\equiv0$ in Eq.~(\ref{Eq:DetailedCoeff})). Due to relativistic effects, the sensitivity quickly grows with the nuclear charge~\cite{FlaDzu09}: for ion clocks $K_\alpha = 8\times 10^{-3}$ for Al$^+$  and $-3$ for Hg$^+$. For lattice clocks, $K_\alpha = 6 \times 10^{-2}$ for Sr, $0.3$ for Yb,
and $0.8$ for Hg.   Then, for example, for  Sr optical lattice clocks, the sum $S$, Eq.~(\ref{Eq:sumS}), simplifies to a single term
\[
S(\mathrm{Sr}) =  \frac{ 6 \times 10^{-2}}{\Lambda_\alpha^2} \,.
\]

Apparently to fully determine energy scales $\Lambda_\alpha$,  $\Lambda_q$,  and $\Lambda_{me/p}$ one would in general require at each node at least two microwave clocks and one optical clock or three microwave clocks. To probe the sensitivity to $\Lambda_\alpha$ a single type of optical clock  populating network nodes is sufficient. It worth noting that as we track transient variation of fundamental constants, it is sufficient to have identical clocks on different nodes. This differs from the search for a slow-drift-in-time of fundamental constants where a typical experiment (see, e.g., Ref.~\cite{RosHumSch08}) uses two co-located clocks with different sensitivity coefficients.


Signal-to-noise ratio~(\ref{Eq:clockLimit}) is affected not only by sensitivity coefficients  but  also  by clock instabilities and sampling rates.
From Eq.(\ref{Eq:clockLimit}), the relevant figure of merit which depends only on the clock parameters (for a given fundamental constant $X$) is 
\begin{equation}
\mathcal{F}=  \frac{ K_X}{ T^{3/2} \sigma_y(T)} \, . \label{Eq:FOM}
\end{equation}

The noise in Eq.~(\ref{Eq:FOM}) scales with the time interval between successive clock comparisons as $T^{3/2} \sigma_y(T)$. 
For atomic clocks, the Allan variance is a complicated function of $T$, depending on which noise source is dominant~\cite{BarChiCut71,Lev99}. Clock instability may even increase with time  but it typically scales down as $\sigma_y(T) \propto 1/\sqrt T$.  Then $\mathcal{F} \propto 1/T $ and  it is beneficial to work with shorter measurement intervals.  The minimum time between consecutive measurements is determined by several factors: in lattice clocks~\cite{DerKat11}, this would be an atomic ensemble preparation  time  (about 1 second), in microwave  fountain clocks~\cite{WynWey05} it the time of flight across interrogation chamber (also about 1 second).  Typically, $\sigma_y(1 \, \mathrm{s} ) \sim 10^{-15}$; then $T^{3/2} \sigma_y(T) = 10^{-15} \, \mathrm{s}^{3/2}$.  Shorter measurement times can be used with active hydrogen masers, where 
for  short time intervals,  $\sigma_y(T) \sim 1/T$, so that  $T^{3/2} \sigma_y(T) \sim \sqrt T$.
For a millisecond time the hydrogen maser exhibits instability of $10^{-10}$~\cite{AllAshHod97}, so that $T^{3/2} \sigma_y(T) \approx 3 \times 10^{-15}$, i.e.,  comparable to the best optical clocks.

Optical lattice clocks \cite{DerKat11} may be  best suited for TDM search due to their stability and accuracy. As of now, Sr lattice clock~\cite{BloNicWil14} has demonstrated the record accuracy at $6 \times 10^{-18}$. The integration time of 3000 s was vastly shorter than that for similarly-accurate ion clocks. 
Similar advances have been reported for Yb clock~\cite{HinShePhi13}.
The statistical advantage of lattice clocks comes from very large number of atoms being interrogated at the same time. The short-term stability is limited by the quality of the local oscillator. With continuing technological improvements, one may anticipate reaching $\sigma_y(1 \, \mathrm{s}) \sim 10^{-18}$. 

Suppose the TDM drives $\alpha$. Then for Hg and Yb lattice clock the figure of merit~(\ref{Eq:FOM}),  $\mathcal{F}( 1\, \mathrm{s}, \mathrm{Hg/Yb}, \alpha) = 10^{18}$, three orders of magnitude better than that for H-masers, $\mathcal{F}( 1 \, \mathrm{ms}, \mathrm{H}, \alpha) = 10^{15}$. H-masers can still be useful because of their sensitivity to other fundamental constants beyond $\alpha$, Eq. (\ref{Eq:H-sensitivities}) and  ubiquity of this mature technology.


\end{document}